\documentstyle[12pt,epsf]{article}
\begin{document}
\begin{flushright}
{Preprint IHEP 97-2}\\
January 1997\\
\end{flushright}
\begin{center}
{\large\bf $B_c$ MESON at LHC}\\
\vspace*{3mm}
{A.V.Berezhnoy, V.V.Kiselev, A.K.Likhoded, A.I.Onischenko}\\
\vspace*{3mm}
Institute for High Energy Physics\\
Protvino, Moscow Region, 142284, Russia.
\end{center}

\begin{abstract}
{In the framework of perturbative QCD and potential models of heavy quarkonium,
total cross-sections and differential characteristics of hadronic
production for different spin states of the $B_c$-meson family are calculated
at energies of LHC. Theoretical predictions for 
masses and branching ratios of $B_c$ decays are given, which allows one to 
make estimates for an expected number of reconstructed events with $B_c$.} 
\end{abstract}

\section*{Introduction}
Along with a possible direct observation of higgs particles and new particle
interactions beyond the Standard Model, a detail investigation
of processes with $b$-quarks, involving mechanisms caused by high orders of
the QCD perturbation theory, is one of the most important tasks of LHC physics
program. Certainly, in the hadron decays with $b$-quarks
the precision study of electroweak model parameters and the observation of 
"new" physics effects are accessible. The latters appear as virtual 
corrections to the leading approximation and, thus, must be reliably separated 
from the QCD high order corrections. So, the investigation of 
$CP$-violation in the heavy quark sector supposes an observation of more than 
$10^9$ decays of $B$ mesons. While performing this
experiment it is also possible to solve another problem such as
the observation of $B_c$ meson and investigation of its properties, which are
attracting a significant interest \cite{ufn}.
With respect to the cross-section for the production of $B$ mesons, the yield
of $B_c$ is suppressed by a hard production of the additional heavy $c$-quark
as well as a small probability of the $(\bar b c)$ state formation, so that
with $\sigma(B_c^{\pm})/\sigma(b\bar b)\sim 10^{-3}$ one can expect about
$10^6$ events with $B_c$ per $10^9$ $B$-mesons.

The $(\bar b c)$ system is a mixed flavour heavy quarkonium, whose
states have no annihilation decay modes due to the electromagnetic and
strong interactions. Therefore, the excited levels of $(\bar b c)$, lying 
below the threshold of decay into the pair of heavy $B$ and $D$ mesons, 
follow the radiation decays into the pseudoscalar ground state,
$B_c^+$ meson, which is a longlived particle, decaying due to the weak 
interaction, $\tau(B_c)\simeq 0.55\pm 0.15$ psec. A nonrelativistic motion 
of heavy quarks in the $(\bar b c)$ system causes a reliable application of 
phenomenological potential models worked out for the charmonium $(\bar c c)$ 
and bottomonium $(\bar b b)$, where the precision of level mass prediction 
reaches 30 íeV \cite{ger,eq}. Weak exclusive decays of $B_c$ are described in 
the framework of quark models for form-factors of different currents 
\cite{sem,lus}, which allows one 
also to determine the total width of this state 
in the form of a sum over all possible decay modes calculated. 
The value obtained in this way agrees with estimates of inclusive $B_c$ 
decays described in the framework of the quark-hadron duality
approach and operator product expansion \cite{ben}. In this consideration one
takes into account the virtualities of heavy quarks in the initial state,
annihilation decay channels and Pauli interference effects for the $c$-quark,
being a product of the $b$-quark decay mode, with the $c$-quark from
the initial $(\bar b c)$ state. 

The $B_c$ production mechanisms can be reliably investigated in the 
framework of perturbative QCD for a hard production 
of two heavy quark pairs and a potential model of a soft nonperturbative 
binding of $\bar b$ and $c$ quarks 
with a small relative momentum in the system of quark mass centre.  
In $e^+e^-$-annihilation at large energies ($M^2_{B_c}/s\ll 1$),
the consideration of leading diagrams for the $B_c$-meson production gives 
the factorized scaling result for the differential cross-section over  
the energy fraction carried out by meson,
$d\sigma/dz= \sigma(\bar b b)\cdot D(z)$, where $z=2E_{B_c}/\sqrt{s}$ 
\cite{ann,frag}. This distribution allows the interpretation as
the hard production of heavier $\bar b$-quark with the subsequent fragmentation
into $B_c$, so that $D(z)$ is the fragmentation function. In this approach
one can obtain analytic expressions for the fragmentation functions into
the different spin states of $S$, $P$, and $D$-wave levels \cite{frag}. 
In photon-photon interactions for the $B_c$ production in the leading
approximation of perturbation theory \cite{phot}, one can isolate three
gauge-invariant groups of diagrams, which can be interpreted as 

1) the hard photon-photon production of $b\bar b$ with 
the subsequent fragmentation of $\bar b \to B_c^+(nL)$, 
where $n$ is the principal quantum number of the $(\bar b c)$-quarkonium, 
$L=0,1\ldots$ is the orbital angular momentum, 

2) the corresponding production and fragmentation for the $c$-quarks, and 

3) the recombination diagrams of $(\bar b c)$-pair
into $B_c^+$, wherein the quarks of different flavours are 
connected to the different photon lines.

In this case, the results of calculation for the complete set of diagrams
in the leading order of perturbation theory show that the group of
$b$-fragmentation diagrams at high transverse momenta
$p_T(B_c)\gg M_{B_c}$ can be described by the fragmentation model with the
fragmentation function $D_{\bar b\to B_c^+}(z)$, calculated in the 
$e^+e^-$-annihilation. The set of $c$-fragmentation diagrams does not
allow the description in the framework of fragmentation model. The 
recombination diagrams give the dominant contribution to the total 
cross-section for the photon-photon production of $B_c$. 

At the LHC energies, the parton subprocess of gluon-gluon fusion 
$gg\to B_c^+ +b +\bar c$ dominates in the hadron-hadron production of 
$B_c$ mesons. In the leading approximation of QCD perturbation theory it 
requires  the calculation of 36 diagrams in the fourth order over the 
$\alpha_s$ coupling constant. In this case, there are no
isolated gauge-invariant groups of diagrams, which would allow the 
interpretation similar to the consideration of $B_c$ production in 
$e^+e^-$-annihilation and photon-photon collisions. 

The calculation for the complete set of diagrams of the
$O(\alpha_s^4)$-contribution
allows one to determine a value of the transverse momentum $p_T^{min}$, 
being the low boundary of the region, where
the subprocess of gluon-gluon $B_c$-meson production enters 
the regime of factorization for the hard production $b\bar b$-pair and 
the subsequent fragmentation of $\bar b$-quark into the bound 
$(\bar b c)$-state, as it follows from the theorem on the
factorization of the hard processes in the perturbative QCD \cite{hadr}. 

The $p_T^{min}$ value turns out to be much greater than the $M_{B_c}$ mass, 
so that the dominant contribution into the total cross-section of
gluon-gluon $B_c$-production is given by the diagrams of 
nonfragmentational type, i.e. by the recombination of heavy quarks. 
Furthermore, the convolution of 
the parton cross-section with the gluon distributions inside the initial 
hadrons leads to the suppression of contributions at large transverse 
momenta as well as the subprocesses with large energy in the system of parton 
mass centre, so that the main contribution into the total cross-section of 
hadronic $B_c$-production is given by the region of energies less or
comparable to the $B_c$-meson mass, where the fragmentation model can not be 
applied by its construction. Therefore, one must perform the calculations 
with the account for all contributions in the given order under consideration
in the region near the threshold.

In this work we calculate the both total and differential cross-sections
for the production of $S$- and $P$-wave states of $B_c$ mesons in the 
framework of leading order of the QCD perturbation theory. Below we give
a quite complete set of the $B_c$ characteristics. The
basic spectroscopic characteristics of the $(\bar b c)$ family and the 
branching ratios of different decay modes of the pseudoscalar ground state, 
which can be used for the extraction of the $B_c$ signal from the hadronic 
background, are presented in section 1. The calculations of 
characteristics in the hadronic $B_c$ production are analyzed in section 2. 
In Conclusion the obtained results are summarized.

\section{The spectroscopy and decays of $B_c$}

\subsection{Level masses and coupling constants}
As a system containing two heavy quarks, the $(\bar b c)$-mesons can be
reliably described in the framework of phenomenological nonrelativistic
potential models \cite{pot} and in QCD sum rules \cite{sr}, which are
based on (i) the formalism of the expansion over the inverse heavy quarks mass,
due to a small ratio of the typical quark virtualities, being of the order of
the confinement scale, to the mass, and (ii) 
the expansion over a small relative 
velocity of the quark motion in the bound state. The perturbative QCD 
allows one to take into account hard gluon corrections to the currents 
of effective quarks, defining the leading approximation in these 
expansions \cite{bra}. This way of consideration leads to a quite precise 
and reliable description of the families of $(\bar c c)$-charmonium and 
$(\bar b b)$-bottomonium, whose spectroscopic properties are in detail 
studied experimentally \cite{pdg}.

The effective gluon potential of static quarks does not depend on the
flavour of heavy quark. The different phenomenological models lead to the 
potential, whose form in the range of average distances between the quarks for
ground and excited levels: $0.1 < r < 1$ fm, really does not depend 
on the model \cite{eic}. The experimental data on the $(\bar c c)$ 
and $(\bar b b)$
systems point out the energy level regularity expressed in the 
approximate quark-flavour independence of the difference between energies of
excitations. According to the Hell-Mann -- Feynman theorem, this regularity is 
caused by the constant value of average kinetic energy of the heavy quarks.
This energy does not depend on the quark flavours and number of excitation. 
The independence becomes exact for the logarithmic potential \cite{log}. 
The weak dependence of the difference between energy levels can be taken into 
account while considering Martin potential: $V(r)=A (r/r_0)^a + C$ \cite{mart}.

The mass values for the different states of 
$(\bar b c)$ system in Martin potential with account for the 
spin-dependent splitting of $nL$-levels are presented in table 1.
The splittings are calculated in the 
second order of $1/m_Q$-expansion over the inverse heavy quark mass with 
the account for the vector coupling of the effective one-gluon exchange 
between the quarks and scalar coupling for the part of potential confining 
the quarks in the bound state \cite{fein}. The effective value of the 
gluon coupling constant at the scale of the average momentum transfer 
between the quarks can be determined 
over the known experimental value for the splitting of charmonium $1S$-level
with the taking into account the renormalization group evolution 
towards the $(\bar b c)$ energies \cite{ger}. 

\begin{table}[t]
\caption{The masses of bound $(\bar b c)$-states below the threshold of decay
into the pair of heavy mesons $BD$, in GeV, in the model with Martin potential
and  the $BT$-potential motivated by QCD with the account for the two-loop 
evolution of the coupling constant at short distances [19]. The spectroscopic 
notations of states are $n^{2j_c}L_J$, where $j_c$ is the total angular 
momentum of $c$-quark, $n$ is the principal quantum number, $L$ is the 
orbital angular momentum, $J$ is the total spin of meson.}
\label{t1}
\begin{center}
\begin{tabular}{||l|c|c||}
\hline
state & Martin\cite{ger}  & $BT$\cite{eq} \\
\hline
$1^1S_0$    & 6.253       & 6.264    \\
$1^1S_1$    & 6.317       & 6.337    \\
$2^1S_0$    & 6.867       & 6.856    \\
$2^1S_1$    & 6.902       & 6.899    \\
$2^1P_0$    & 6.683       & 6.700    \\
$2P\; 1^+$  & 6.717       & 6.730    \\
$2P\; 1'^+$ & 6.729       & 6.736    \\
$2^3P_2$    & 6.743       & 6.747    \\
$3^1P_0$    & 7.088       & 7.108    \\
$3P\; 1^+$  & 7.113       & 7.135    \\
$3P\; 1'^+$ & 7.124       & 7.142    \\
$3^3P_2$    & 7.134       & 7.153    \\
$3D\; 2^-$  & 7.001       & 7.009    \\
$3^5D_3$    & 7.007       & 7.005    \\
$3^3D_1$    & 7.008       & 7.012    \\
$3D\; 2'^-$ & 7.016       & 7.012    \\
\hline
\end{tabular}
\end{center}
\end{table}

Thus, in the $(\bar b c)$ system below the threshold of decay into the 
heavy meson $BD$ pair, one can expect the presence of 16 narrow states of the
$1S$, $2S$,$2P$, $3P$ and $3D$-levels, which cascadely transform into the 
ground pseudoscalar state with the mass $m_{B_c}=6.25\pm 0.03$ GeV due to the 
emission of photons and $\pi$-meson pairs, since the annihilation 
$(\bar b c)$ decays can occur only due to the weak interaction and, hence, are 
suppressed for the excited levels. The total widths of the excited states and 
the branching ratios for the radiation modes of decay are shown in table 2.

The potentials motivated by QCD have the linear growth at large distances
and the coulumb-like behaviour at short ones. The form of the potential 
in the region, where the perturbative regime changes into the 
nonperturbative one with the increase of
the distance between the quarks, coincides with the form of the logarithmic or
power potentials, so that the accuracy for the prediction of the energy levels
in the heavy quarkonia, particular in the $(\bar b c)$ system, is determined by
the value of 30 MeV. 

The global properties of these potentials, i.e. their asymptotic behaviour
in the bound points of $r\to \infty$, $r\to 0$, are significant for the 
determination of the coupling constants of the states, such as the leptonic 
coupling constants $f$ for the $nS$-levels, for example. In the leading 
approximation, the $f$ value does not depend on the spin state of the level 
and it is determined by the value of the radial wave function at the origin, 
$R(0)$,
$$
\tilde f_{n} = \sqrt{\frac{3}{\pi M_{n}}} R_{nS}(0)\;.
$$
Taking into account the hard gluon corrections \cite{f0,scale}, the constants 
of vector and pseudoscalar states equal
\begin{equation}
f_{n}^{V,P} = \tilde f_{n}\; \biggl(1+\frac{\alpha_s}{\pi}
\biggl(\frac{m_1-m_2}{m_1+m_2}\ln\frac{m_1}{m_2}-\delta^{V,P}\biggr)\biggr)\;,
\end{equation}
where $m_{1,2}$ are the quark masses, $\delta^V=8/3$, $\delta^P=2$,  and
the QCD coupling constant is determined at the scale of the gluon virtuality,
given by the quark masses function \cite{vol}.
\begin{table}[t]
\caption{The total widths of excited bound $(\bar b c)$-states below the
threshold of decay into the meson $BD$-pair in the model with Martin potential
and the branching ratios of the dominant decay modes.}
\label{t2}
\begin{center}
\begin{tabular}{||l|r|p{29.3mm}|r||}
\hline
state & $\Gamma_{\rm tot}$, KeV  &dominant decay mode & BR, \% \\
\hline
$1^1S_1$    & 0.06       & $1^1S_0+\gamma$   & 100  \\
$2^1S_0$    & 67.8       & $1^1S_0+\pi\pi$   &  74  \\
$2^1S_1$    & 86.3       & $1^1S_1+\pi\pi$   &  58  \\
$2^1P_0$    & 65.3       & $1^1S_1+\gamma$   & 100  \\
$2P\; 1^+$  & 89.4       & $1^1S_1+\gamma$   &  87  \\
$2P\; 1'^+$ & 139.2      & $1^1S_0+\gamma$   &  94  \\
$2^3P_2$    & 102.9      & $1^1S_1+\gamma$   & 100  \\
$3^1P_0$    & 44.8       & $2^1S_1+\gamma$   &  57  \\
$3P\; 1^+$  & 65.3       & $2^1S_1+\gamma$   &  49  \\
$3P\; 1'^+$ & 92.8       & $2^1S_0+\gamma$   &  63  \\
$3^3P_2$    & 71.6       & $2^1S_1+\gamma$   &  69  \\
$3D\; 2^-$  & 95.0       & $2P\; 1^++\gamma$ &  47  \\
$3^5D_3$    & 107.9      & $2^3P_2+\gamma$   &  71  \\
$3^3D_1$    & 155.4      & $2^1P_0+\gamma$   &  51  \\
$3D\; 2'^-$ & 122.0      & $2P\; 1'^++\gamma$ & 38 \\
\hline
\end{tabular}
\end{center}
\end{table}

The values of the heavy quarkonium wave functions are significant in the
consideration of production and decays of their states in the framework
of quark models of the mesons. They are shown in table 3 for $S$- and 
$P$-wave states in the model with Martin potential. 
However, as it was mentioned above, the accuracy of potential models
for the values under consideration is quite low, and the uncertainty is expressed by
a factor of two.
\begin{table}[t]
\caption{The wave function characteristics, $R_{nS}(0)$ (in GeV$^{3/2}$)
and $R'_{nP}(0)$ (in GeV$^{5/2}$), obtained from Schr\" odinger equation
with Martin potential and $BT$.}
\label{t3}
\begin{center}
\begin{tabular}{||c|c|c||}
\hline
n & Martin\cite{ger}  & $BT$\cite{eq}\\
\hline
$R_{1S}(0)$  & 1.31 & 1.28 \\
$R_{2S}(0)$  & 0.97 & 0.99 \\
$R'_{2P}(0)$ & 0.55 & 0.45 \\
$R'_{3P}(0)$ & 0.57 & 0.51 \\
\hline
\end{tabular}
\end{center}
\end{table}

The QCD sum rules allow one to determine the leptonic constants for the heavy
quarkonium states with a much better accuracy. Standard schemes of the sum 
rules give an opportunity to calculate the ground state constants for 
vector and pseudoscalar currents with the account for corrections
from the quark-gluon condensates, which have the power form over the inverse
heavy quark mass \cite{sr}. There is a region of the momentum numbers
for the spectral density of the two-point current correlator, where the
condensate contributions are not significant. In this region,
the integral representation for the contribution
of resonances lying bellow the threshold of decay into the pair of
heavy mesons, allows one to use the regularity of the quarkonium state density
mentioned above, and to derive the scaling relations for the leptonic 
constants of the ground state quarkonia with different quark contents and 
for the excited states \cite{scale}. 
Thus, for vector states we have
\begin{equation}
\frac{f^2_n}{M_n}\;\biggl(\frac{M_n}{M_1}\biggr)^2\;
\biggl(\frac{m_1+m_2}{4m_{12}}\biggr)^2 = \frac{c}{n}\;,
\label{scal}
\end{equation}
where $m_{12}=m_1m_2/(m_1+m_2)$ is the reduced mass of quarks,
and the constant $c$ is determined by the average kinetic energy of quarks,
the QCD coupling constant at the scale of average momentum transfers
in the system and the hard gluon correction factor to the vector current.
Numerically, in the method accuracy the $c$ value turns out to depend on
no quark flavour and excitation number in the system. Relation (\ref{scal})
is in a good agreement with  the data on the coupling constants for
the families of $\psi$ and $\Upsilon$ particles and, thus, it can be
a reliable basis for the prediction of leptonic constants for the
$B_c$-meson family, as it is shown in table 4, where one can also find
the constants for the pseudoscalar states of $nS$-levels, which are
determined by the relation (see \cite{scale})
$$
f_n^P = f_n \biggl(1+\frac{2\alpha_s}{3\pi}\biggr)\; \frac{m_1+m_2}{M_n}\;.
$$

The $R(0)$ values, being the parameters of
static quark models in the consideration of decays and production of 
the mesons, can be determined over the leptonic constants calculated in
the sum rules. In this way, the additional uncertainty appears. It is
related with a choice of the scale fixing the QCD coupling constant in
the factor accounting for the hard gluon correction. The dependence 
on this choice points out the significance of high orders in the perturbative 
approximation. Assuming that in the $\overline{\rm MS}$-scheme the scale is 
close to $\mu^2=m_1m_2e^{-11/12}$ \cite{vol}, one obtains the values of 
$R(0)$ shown in table 4, where one can see that the predictions of 
sum rules and the potential models of Martin and $BT$ are in a reasonable 
agreement with each other. 
\begin{table}[t]
\caption{The leptonic constants for the vector and pseudoscalar states of
$nS$-levels in the $(\bar b c)$-system, $f_n$ and $f_n^P$, calculated in the
sum rules resulting in the scaling relation, and $R_{nS}(0)$, calculated over $f_n$. 
The accuracy is equal to 6\%.}
\label{t4}
\begin{center}
\begin{tabular}{||c|c|c|c||}
\hline
n & $f_n$, MeV  & $f_n^P$, MeV & $R_{nS}(0)$, GeV$^{3/2}$ \\
\hline
1 & 385 & 397 & 1.20 \\
2 & 260 & 245 & 0.85 \\
\hline
\end{tabular}
\end{center}
\end{table}

As for the states lying above the threshold of decay into the heavy meson 
$BD$ pair, the width of $B_c^{*+}(3S)\to B^+D^0$, for example, can be calculated
in the framework of sum rules for the meson currents, where the scaling relation
takes place for the $g$ constants of similar decays of
$\Upsilon(4S)\to B^+B^-$ and $\psi(3770)\to D^+D^-$ \cite{g-sr},
$$
\frac{g^2}{M}\; \biggl(\frac{4m_{12}}{M}\biggl)= const.
$$
The relation is caused by the dependence of energy gap between the vector and 
pseudoscalar heavy meson states: $\Delta M_{1,2}\cdot M_{1,2} = const.$,
where $M_{1,2}$ are the meson masses in the final state, $m_{12}$ is
their reduced mass. The width of this decay has a strong dependence
on the $B_c^{*+}(3S)$ mass, and at $M=7.25$ GeV it is equal to 
$\Gamma= 90\pm 35$ MeV, where the uncertainty is determined by the accuracy of 
method used.

\subsection{The $B_c$ meson decays}
\subsubsection{Life-time}
The $B_c$-meson decay processes can be subdivided into three classes:
1) the $\bar b$-quark decay with the spectator $c$-quark, 2) the 
$c$-quark decay 
with the spectator $\bar b$-quark and 3) the annihilation channel
$B_c^+\rightarrow l^+\nu_l (c\bar s, u\bar s)$, where $l=e,\; \mu,\; \tau$.
In the $\bar b \to \bar c c\bar s$ decays one separates 
also the Pauli interference 
with the $c$-quark from the initial state.
In accordance with the given classification, 
the total width is the sum over the partial widths
\begin{equation}
\Gamma (B_c\rightarrow X)=\Gamma (b\rightarrow X)
+\Gamma (c\rightarrow X)+\Gamma \mbox{(ann.)}+\Gamma\mbox{(PI)}\;.
\end{equation}
For the annihilation channel the  $\Gamma\mbox{(ann.)}$ width can
be reliably estimated in the framework of inclusive approach, where one
take the sum of the leptonic and quark decay modes with account for the hard
gluon corrections to the effective four-quark interaction of weak
currents. These corrections result in the factor of 
$a_1=1.22\pm 0.04$ \cite{ufn}. The width is expressed through the
leptonic constant of $f_{B_c}=f_1^P\approx 400$ MeV. This estimate of the 
quark-contribution does not depend on a hadronization model, since the 
large energy release of the order of the meson mass takes place.
From the following expression one can see that one can neglect the contribution
by light leptons and quarks,
\begin{equation}
\Gamma \mbox{(ann.)} =\sum_{i=\tau,c}\frac{G^2_F}{8\pi}
|V_{bc}|^2f^2_{B_c}M m^2_i (1-m^2_i/m^2_{Bc})^2\cdot C_i\;,
\label{d3}
\end{equation}
where $C_\tau = 1$ for the $\tau^+\nu_\tau$-channel and 
$C_Ó =3|V_{cs}|^2a_1^2 $ for the $c\bar s$-channel.

As for the nonannihilation decays, in the approach of the operator 
product expansion for the quark currents of weak decays \cite{ben},
one takes into account the $\alpha_s$-corrections to the free quark decays and
uses the quark-hadron duality for the final states. Then one considers the 
matrix element for the transition operator over the bound meson state.
The latter allows one also to
take into account effects caused by the motion and virtuality of decaying 
quark inside the meson because of the interaction with the spectator.
In this way the $\bar b\to \bar c c\bar s$ decay mode turns to be 
suppressed almost completely due to the Pauli interference with the charm quark
from the initial state. Besides, the $c$-quark decays with the 
spectator $\bar b$-quark is essentially suppressed in comparison with
the free quark decays because of a large bound energy in the initial state. 

In the framework of exclusive approach it is necessary to sum widths 
of different decay modes calculated in the potential models \cite{sem,lus}. 
While considering the semileptonic decays due to the
$\bar b \to \bar c l^+\nu_l$ and $c\to s l^+\nu_l$ transitions
one finds that  in the former decays the hadronic final state is practically
saturated by the lightest bound $1S$-state in the $(\bar c c)$-system, 
i.e. by the $\eta_c$ and $J/\psi$ particles, and in the latter decays
the $1S$-states in the $(\bar b s)$-system, i.e. $B_s$ and 
$B_s^*$, only can enter the accessible energetic gap. 
The energy release in the latter transition is low in comparison with 
the meson masses, and, therefore, a visible deviation from the picture
of quark-hadron duality is possible. Numerical estimates show that the
value of $B_c\to(\bar b s)l^+\nu_l$ decay width is two times less 
in the exclusive approach than in the inclusive method, though this fact can
be caused by the choice of narrow wave package for the $B_s^{(*)}$ mesons in 
the quark model, so that $\tilde f_{B_s}\approx 150$ MeV, while
in the limit of static heavy quark, one should expect a larger value for the
leptonic constant \cite{neu,fmu}. This increase will lead to the widening
of the wave package and, hence, to the increase of the overlapping integral
for the wave functions of $B_c$ and $B_s^{(*)}$ (see table 5).
\begin{table}[t]
\caption{The branching ratios of the $B_c$ decay modes calculated in
the framework of inclusive approach and in the exclusive quark model with 
the parameters $|V_{bc}|=0.040$, $\tilde f_{B_c}=0.47$ GeV, 
$\tilde f_{\psi}=0.54$
GeV,$\tilde f_{B_s}=0.3$ GeV, $m_b=4.8-4.9$ GeV, $m_c=1.5-1.6$ GeV,
$m_s=0.55$ GeV. The accuracy is about 10\%.}
\label{t5}
\begin{center}
\begin{tabular}{||l|c|c||}
\hline
$B_c$ decay mode & Inclus., \%  & Exclus., \% \\
\hline
$\bar b\to \bar c l^+\nu_l$ & 3.9 & 3.7 \\
$\bar b\to \bar c u\bar d$  & 16.2 & 16.7 \\
$\sum \bar b\to \bar c$     & 25.0 & 25.0 \\
$c\to s l^+\nu_l$           & 8.5  & 10.1 \\
$c\to s u\bar d$            & 47.3 & 45.4 \\
$\sum c\to s$               & 64.3 & 65.6 \\
$B_c^+\to \tau^+\nu_\tau$   & 2.9  & 2.0  \\
$B_c^+\to c\bar s$          & 7.2  & 7.2  \\
\hline
\end{tabular}
\end{center}
\end{table}

Further, the $\bar b\to \bar c u\bar d$ channel, for example, can be calculated
through the given decay width of $\bar b \to \bar c l^+\nu_l$ with account
for the color factor and hard gluon corrections to the four-quark 
interaction. It can be also obtained as a sum over the widths of decays 
with the $(u\bar d)$-system bound states. 

The results of calculation for the total $B_c$ width in the inclusive and
exclusive approaches give the values consistent with each other, if one 
takes into account the most significant uncertainty related with the choice
of quark masses (especially for the charm quark), so that finally we have 
\begin{equation}
\tau(B_c^+)= 0.55\pm 0.15\; \mbox{psec.}
\end{equation}

\subsubsection{Exclusive decays}
The consideration of exclusive $B_c$-decay modes supposes an introduction
of model for the hadronization of quarks into the mesons with the
given quantum numbers. The QCD sum rules for the three-point correlators of 
quark currents \cite{pav,lep-sr} and potential models \cite{sem,lus} are 
among of those hadronization models. A feature of the sum rule application
to the mesons containing two heavy quarks, is the account for the significant 
role of the coulumb-like $\alpha_s/v$-corrections due to the gluon exchange 
between the quarks composing the meson and moving with the relative velocity
$v$. So, in the semileptonic decays of $B_c^+\to \psi(\eta_c)l^+\nu_l$, 
the heavy $(\bar Q_1Q_2)$ quarkonium is present in the both initial and 
final states, and, therefore, the contribution of coulumb-like corrections 
exhibits in a specially strong form. The use of tree approximation for 
the perturbative contribution into the three-point correlator of quark 
currents \cite{pav} leads to a large deviation between the values of 
transition form-factors, calculated in the sum rules and 
potential models, respectively. The account for the $\alpha_s/v$-corrections 
removes this contradiction \cite{lep-sr}.
Thus, the meson potential models, based on the covariant expression for the
form-factors of weak $B_c$ decays through the overlapping of quarkonium wave 
functions in the initial and final states, and the QCD sum rules give 
the consistent description of semileptonic $B_c$-meson decays.  

Further, the hadronic decay widths can be obtained on the basis of assumption
on the factorization of the weak transition between the quarkonia and 
the hadronization of products of the virtual $W^{*+}$-boson decay \cite{fact}.
The accuracy of factorization has to raise with the increase of $W$-boson 
virtuality. This fact is caused by the suppression of interaction in the 
final state. In this way, the hadronic decays can be calculated due to the
use of form-factors for the semileptonic transitions with the 
relevant description of $W^*$ transition into the hadronic state.

Let us consider the amplitude of $B_c^+ \to M_X e^+ {\nu}_e$ transition 
with the weak decay of quark 1 into the quark 2 
\begin{equation}
A = \frac{G_F}{\sqrt{2}}\;V_{12}\;l_{\mu}\;H^{\mu}\;, 
\label{d12}
\end{equation}
where $G_F$ is the Fermi constant, $V_{12}$ is the Kobayashi-Maskawa 
matrix element. The leptonic $l_{\mu}$ current is determined by the expression
\begin{equation}
l_{\mu} = \bar e(q_1)\gamma_{\mu} (1 - \gamma_5) \nu(q_2)\;,
\end{equation}
where $q_{1,2}$ are the momenta of lepton and neutrino, correspondingly;
$(q_1+q_2)^2 = t$.
The $H_\mu$ value in (\ref{d12}) is the matrix element for the hadronic
$J_\mu$ current 
\begin{equation}
J_{\mu} = V_\mu - A_\mu = \bar Q_1 \gamma_{\mu} (1 - \gamma_5) Q_2\;.
\end{equation}
The matrix element for the $B_c$ decay into the pseudoscalar $P$ state can be 
written down in the form
\begin{equation}
<B_c(p)| A_{\mu} |P(k)> = F_+(t) (p+k)_{\mu} + F_-(t) (p-k)_{\mu}\;, 
\label{d16}
\end{equation}
and for the transition into the vector $V$ meson with the $M_V$ mass 
and $\lambda$ polarization one can express the matrix element in the form
\begin{eqnarray}
<B_c(p)| J_{\mu} |V(k,\lambda)> & = & 
-(M+M_V) A_1(t)\;\epsilon_{\mu}^{(\lambda)} \nonumber \\
& ~ & +  \frac{A_2(t)}{M+M_V}\;(\epsilon^{(\lambda)} p)\;
(p+k)_{\mu} \nonumber \\
& ~ &  + \frac{A_3(t)}{M+M_V}\;(\epsilon^{(\lambda)} p)\;
(p-k)_{\mu}  \nonumber \\
& ~ &  + i \frac{2 V(t)}{M+M_V}\;\epsilon_{\mu \nu \alpha \beta}\;
\epsilon^{(\lambda)\nu} p^\alpha k^\beta.
\label{d17}
\end{eqnarray}
Relations (\ref{d16}), (\ref{d17}) define the form-factors for the 
$B_c^+ \to M_X e^+ {\nu}_e$ transitions, so that for the massless leptons, 
$F_-$ and $A_3$ do not contribute into the matrix element in (\ref{d12}).
In the covariant quarkonium model with the wave functions of oscillator type
\cite{sem} one can easily obtain
\begin{eqnarray}
F_+(t) & = & \frac{1}{2}\; (m_1+m_2) \sqrt{\frac{M_P}{M}} \frac{1}{m_2} 
\xi _P (t)\;,\\ \label{d18}
F_-(t) & = & -\frac{1}{2}\; (m_1-m_2+2 m_{sp}) \sqrt{\frac{M_P}{M}} 
\frac{1}{m_2} \xi _P (t)\;,
\end{eqnarray}
where $m_{sp}$ is the mass of spectator quark, and the $\xi (t)$ function 
has the form
\begin{equation}
\xi _X(t) =  
\frac{2 \omega \omega_X}{\omega^2+\omega_X^2}
\sqrt{\frac{2 \omega \omega_X}{\omega^2 y^2+\omega_X^2}}
\exp\bigg\{-\frac{m_{sp}^2}{\omega^2 y^2+\omega_X^2}\;(y^2-1)\bigg\}\;,
\label{d20}
\end{equation}
where $M_X$ is the meson mass, and $\omega_X$ is the wave function parameter of
the recoil meson,
\begin{equation}
y=1+\frac{t_m-t}{2MM_X},\;\;\;\; t_m = (M-M_X)^2,\;\;\;\;\;
\tilde f_X=\sqrt{\frac{12}{M_X}}\biggl(\frac{\omega_X^3}{2\pi}\biggr)^{3/2}\;,
\end{equation}
$t_m$ is the maximum square of the leptonic pair mass. 
For the vector state $M_V=M_X$ we get
\begin{eqnarray}
V(t) & = & \frac{1}{2}\; (M+M_V) \sqrt{\frac{M_V}{M}} \frac{1}{m_2} \xi _V (t)\;,\label{d22} \\
A_1(t) & = & \frac{1}{2}\; \frac{M^2+M_V^2-t+2 M (m_2-m_{sp})}{M+M_V} \sqrt{\frac{M_V}{M}} \frac{1}{m_2} \xi _V (t)\;,\label{d23}\\
A_2(t) & = & \frac{1}{2}\; (M+M_V) (1-2 m_{sp}/M) \sqrt{\frac{M_V}{M}} \frac{1}{m_2} \xi _V (t)\;,\label{d24}\\
A_3(t) & = & -\frac{1}{2}\; (M+M_V) (1+2 m_{sp}/M) \sqrt{\frac{M_V}{M}} \frac{1}{m_2} \xi _V (t)\;. \label{d25}
\end{eqnarray}
The branching ratios of some exclusive $B_c$-decay modes are
shown in table 6.
\begin{table}[t]
\caption{The branching ratios of exclusive $B_c$ decay modes,
calculated in the framework of covariant quark model with the parameters 
$|V_{bc}|=0.040$, $\tilde f_{B_c}=0.47$ GeV, $\tilde f_{\psi}=0.54$ GeV,
$\tilde f_{B_s}=0.3$ GeV, $m_b=4.8-4.9$ GeV, $m_c=1.5-1.6$ GeV,
$m_s=0.55$ GeV. The accuracy equals 10\%.}
\label{t6}
\begin{center}
\begin{tabular}{||c|c|c|c||}
\hline
$B_c$ decay mode & BR, \%  & $B_c$ decay mode & BR, \% \\
\hline
$\psi l^+\nu_l$ & 2.5 & $\eta_c l^+\nu_l$ & 1.2 \\
$B^*_s l^+\nu_l$ & 6.2 & $B_s l^+\nu_l$    & 3.9 \\
$\psi \pi^+$    & 0.2 & $\eta_c \pi ^+$   & 0.2 \\
$B^*_s \pi^+$    & 5.2 & $B_s \pi^+$       & 5.5 \\
$\psi \rho^+$    & 0.6 & $\eta_c \rho ^+$   & 0.5 \\
$B^*_s \rho^+$    & 22.9 & $B_s \rho^+$       & 11.8 \\
\hline
\end{tabular}
\end{center}
\end{table}

A decrease of the invariant mass for the hadron system  results in an
increase of the recoil meson momentum. This causes the problem of applicability
for the formalism of overlapping for the quarkonium wave functions, 
because, in this kinematics, the narrow wave packages are displaced relative 
to each other in the momentum space into the range of distribution tails.
In this situation one has to take into account a hard gluon exchange
between the quarkonium constituents, which destroys the spectator picture of 
weak transition in the potential approach. For example, the widths of
$B_c^+\to \psi \pi^+$ and $B_c^+\to \eta_c \pi^+$ decays have the 
following forms
\cite{hard}
\begin{eqnarray}
\Gamma(B_c^+\to \psi \pi^+) &=& G_F^2 |V_{bc}|^2\; \frac{128\pi\alpha_s^2}{81}
f^2_\pi \tilde f^2_{B_c}\tilde f^2_{\psi} \biggl(\frac{M+m_\psi}{M-m_\psi}
\biggr)^3\; \frac{M^3}{(M-m_\psi)^2 m_\psi^2}\; a_1^2,
\label{g-}\\
\Gamma(B_c^+\to \eta_c \pi^+) &=& \Gamma(B_c^+\to \psi \pi^+) \cdot
\frac{(5-m^2_{\eta_c}/M^2)^2}{64}\;.
\label{g}
\end{eqnarray}
As one can see in equations (\ref{g-})-(\ref{g}), the basic uncertainty in 
the width estimation is connected with the scale choice for
the "running" QCD coupling constant, which supposes the renormalization 
of the six-quark operator for the transition between the quarkonia in 
the next-to-tree approximation.
This problem requires a carrying out a large volume of cumbersome 
calculations. However, even in this stage of evaluations one can stat 
that the corrections with "large" logarithms expect to appear from 
the contributions related with the renormalization of four-quark weak 
interaction (the $a_1$ factor) and in the gluon propagator standing in 
the exchange between the quarks. The latter leads to the fixing of $\alpha_s$ 
scale at the gluon virtuality. Numerically, this virtuality 
is determined by the recoil meson momentum, and it is slightly greater than 
1 GeV, so that one can talk on a quite hard gluon exchange, which 
gives the values of branching ratios of approximately one order of magnitude
greater than in the potential approach, as one can find in table 7. 
However, the conservative estimate of uncertainty in the hard approximation
is close to 50\%.
\begin{table}[t]
\caption{The branching ratios of exclusive $B_c$ decay modes at the
large momenta of recoil meson in the framework of hard gluon exchange
approximation. The uncertainty equals 50\%.}
\label{t7}
\begin{center}
\begin{tabular}{||c|c|c|c||}
\hline
$B_c$ decay mode & BR., \%  & $B_c$ decay mode & BR, \% \\
\hline
$\psi \pi^+$    & 2.0 & $\eta_c \pi ^+$   & 0.7 \\
$\psi \rho^+$    & 5.6 & $\eta_c \rho ^+$   & 2.0 \\
\hline
\end{tabular}
\end{center}
\end{table}

As for the extraction of $B_c$ signal in the hadronic background, 
the decay modes with $\psi$ in the final state are the most preferable, because
the latter particle can be easily identified by its leptonic decay mode.
This advantage is absent in the $B_c$ decay modes with the final state 
containing the $\eta_c$ or $B^{(*)}_s$ mesons, which are the objects, 
whose experimental registration is impeded by a large hadron background.
From the values of branching ratios shown above one can
easily obtain, that the total probability of $\psi$ yield in the $B_c$ decays
equals ${\rm BR}(B_c^+\to \psi X)=0.24$. It is worth to note that the key
role in the $B_c$ signal observation plays the presence of the vertex
detector, which allows one to extract events with the weak decays of
long-lived particles containing heavy quarks. In the case under consideration
it gives the possibility to suppress the background from the direct $\psi$ 
production. In the semileptonic $B_c^+\to \psi l^+\nu_l$ decays 
the presence of vertex detector and large statistics of events allows one
to determine the $B_c$-meson mass and to separate the events with its decays 
from the ordinary $B_{u}^+$-meson decays, which have no $\psi l^+$ mode. 
In the $\psi\pi^+$ decay the direct measurement of $B_c$ mass is possible.
The detector efficiency in the reconstruction of three-particle secondary
vertex ($l^+l^-$  from decays of $\psi$ and $\pi^+$ or $l^+$) becomes the
most important characteristics here. The low efficiency of the LEP detectors 
($\epsilon\approx 0.15$), for example, makes the $B_c$ observation to be
probably unreachable in the experiments at the electron-positron collider 
\cite{lep}.

\section{The characteristics of $B_c$ production at LHC}
As was mentioned above, the consideration of mechanisms for the hadronic 
production of different spin $B_c$-states is based on the factorization of 
hard parton production of heavy quarks $(\bar b b\bar c c)$ and soft 
coupling of $(\bar b c)$ bound state \cite{hadr}. In the first stage of 
description, the hard subprocess can be reliably calculated in the framework 
of QCD perturbation theory, while in the second stage the quark binding in 
the heavy quarkonium can be described in the nonrelativistic potential model 
assigned to the $(\bar b c)$-pair rest system. The latter means that
one performs the integration of the final quark state over 
the quarkonium wave function in the momentum space. Since the relative 
quark velocity inside the meson is close to zero, the perturbative matrix 
element can be expanded in series over the relative quark momentum, 
which is low in comparison with the quark masses determining the scale 
of virtualities and energies in the matrix element. 
In the leading approximation one considers only the first nonzero 
term of such expansion, so that for the $S$-wave states the 
matrix element of the parton subprocess for the $B_c$ production is expressed 
through the perturbative matrix element for the production of four heavy 
quarks $(gg\to\bar b b \bar c c)$ with the corresponding projection to
the vector or pseudoscalar spin state of $(\bar b c)$-system, 
being the color singlet, and through the factor of radial wave function 
at the origin, $R_{nS}(0)$, for the given quarkonium. 
The perturbative matrix element is calculated for the $(\bar b c)$ state, 
where the quarks move with the same 
velocity, i.e. one neglects the relative motion of $\bar b$ and $c$.

For the $P$-wave states, the potential model gives the factor in the form of 
first derivative of the radial wave function at the origin, $R'_{nL}(0)$.
In the perturbative part, one has to calculate the first derivative of the 
matrix element over the relative quark momentum at the point, where the 
velocities of quarks, entering the quarkonium, equal each to other. 

Thus, in addition to the heavy quark masses, the values of $R_{nS}(0)$, 
$R'_{nL}(0)$ and $\alpha_s$ are the parameters of calculation for the
partonic production of $B_c$ meson.
In calculations we use the wave function parameters equal to the values
shown in table 4 and $R'_{2P}(0)=0.50$ GeV$^{3/2}$.
The value of $R(0)$, as was shown above, can be related with the leptonic 
constant, $\tilde f$, so that we have
$$
\tilde f_{1S} = 0.47\;\mbox{GeV,}\;\;\;
\tilde f_{2S} = 0.32\;\mbox{GeV.}
$$
\begin{figure}[b]
\hspace*{3cm}$d\hat\sigma/dp_T$, nb/GeV\\
\vspace*{-5mm}
\begin{center}
\hspace*{-10mm}
\epsfxsize=9cm \epsfbox{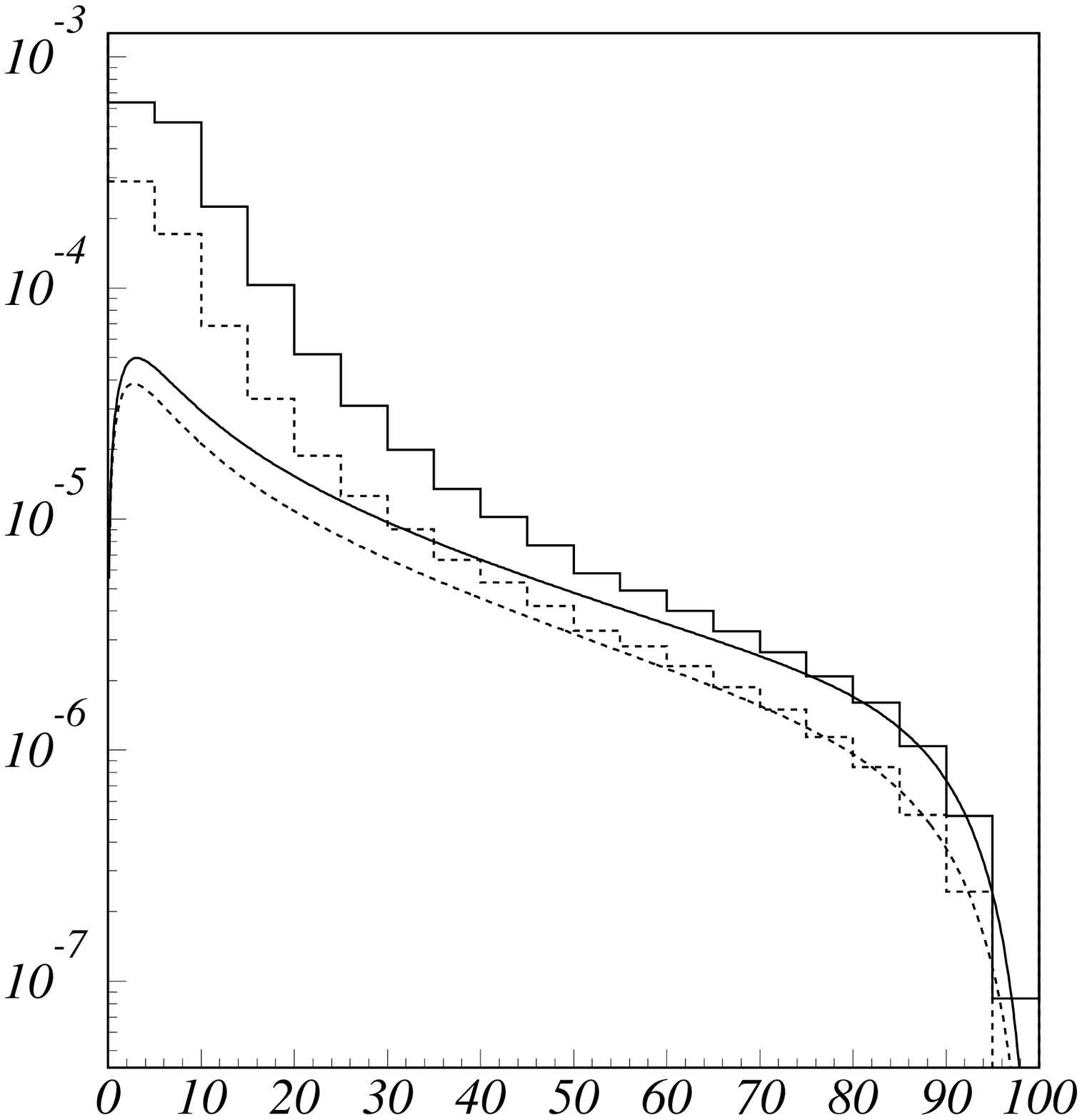}
\end{center}
\vspace*{-13mm}
{\hfill GeV\hspace*{38mm}}
\caption{The differential cross-section for the $B_c^{(*)}$ meson production in
gluon-gluon collisions as calculated in the perturbative QCD over the complete
set of diagrams in the $O(\alpha_s^4)$ order at 200 GeV.
The dashed and solid histograms present the pseudoscalar and vector states, 
respectively, in comparison with the results of fragmentation model
shown by the corresponding smooth curves.}
\label{fig1}
\end{figure}
At large transverse momenta of the $B_c$ meson,  $p_T\gg M_{B_c}$,
the production mechanism enters the regime of $\bar b$-quark fragmentation 
(see fig. 1), so that the scale determining the QCD coupling constant in hard 
$\bar b b$ production, is given by $\mu^2_{\bar b b}\sim M^2_{B_c}+p^2_T$,
and in the hard fragmentation production of the additional pair of heavy quarks
$\bar c c$ we get $\mu_{\bar c c} \sim m_c$. This scale choice is
caused by the high order corrections of perturbation theory to the 
hard gluon propagators, where the summing of logarithms over the virtualities
leads to the pointed $\mu$ values. Therefore, the normalization of matrix 
element is determined by the value of
$\alpha_s(\mu_{\bar b b})\alpha_s(\mu_{\bar c c})\approx 0.18 \cdot 0.28$. 
In calculations we use the single combined value of $\alpha_s=0.22$.

In the approximation of the weak quark binding inside 
the meson one has $M_{B_c}= m_b+m_c$,
so that the performable phase space in calculations is close to physical one
at the choices of
$m_b=4.8$ GeV, $m_c=1.5$ GeV for the $1S$-state,
$m_b=5.1$ GeV, $m_c=1.8$ GeV for the $2S$-state,
$m_b=5.0$ GeV, $m_c=1.7$ GeV for the $2P$-state. 
\begin{figure}[t]
\hspace*{1cm}$d\sigma/dp_T$, nb/GeV \hspace*{3.3cm}~$d\sigma/dp_T$, 
nb/GeV\\
\vspace*{-19mm}
\begin{center}
\hspace*{-10mm}
\epsfxsize=12cm \epsfbox{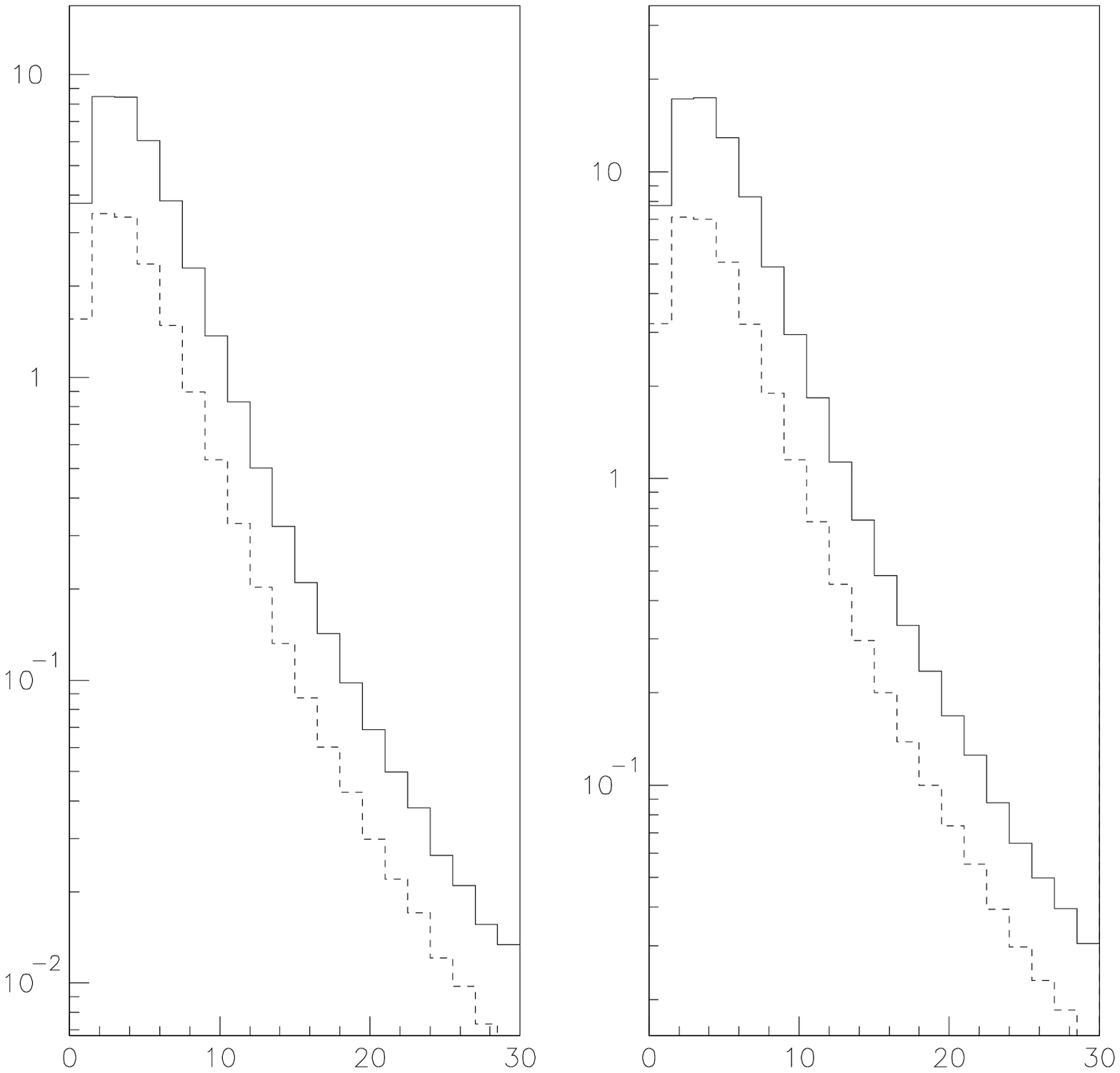}
\end{center}
\vspace*{-29mm}
\hspace*{4cm}a)~~~~~~~~~~$p_T$,~GeV~~~~~~~~~~~~~~~~~~~~~b)~~~~~~~~~$p_T$,~GeV
\caption{The distributions over the transverse momentum of $B_c^*$ 
(solid line) and $B_c$ (dashed line) in proton-proton interactions at the
energies of 8 TeV (a) and 16 TeV (b).}
\end{figure}
The value of the hadron-hadron cross-section is determined by the convolution
of parton distributions with the cross-sections of parton subprocesses.
At the LHC energies the subprocess of $gg\to B_c^+b\bar c$ is dominant
in the production of $B_c$ meson because of the more high gluon-gluon 
luminosity. The total cross-sections of the subprocesses for the different
spin states of $B_c$ meson are calculated numerically, and their dependence
on the subprocess energy, $\sqrt{\hat s}$, can be written down in the following
analytic approximation    
\begin{eqnarray}
\sigma(gg\to B_c(1S)\; b\bar c) & = & 550.0\; 
\biggl(1-\frac{2(m_b+m_c)}{\sqrt{\hat s}}\biggr)^{2.35}\cdot
\biggr(\frac{2(m_b+m_c)}{\sqrt{\hat s}}\biggr)^{1.37}\; \mbox{pb,}\\
\sigma(gg\to B_c(2P)\; b\bar c) & = & 25.0\; 
\biggl(1-\frac{2(m_b+m_c)}{\sqrt{\hat s}}\biggr)^{1.95}\cdot
\biggr(\frac{2(m_b+m_c)}{\sqrt{\hat s}}\biggr)^{1.2}\; \mbox{pb,}
\end{eqnarray}
in the region of $2(m_b+m_c)< \sqrt{\hat s}< 400$ GeV.

The using of CTEQ4M parameterization for the structure functions
of nucleon \cite{cteq} leads to the total hadronic cross-sections 
for the $B_c$ mesons, as shown in table 8. After the summing over the
different spin states, the total cross-sections for
the production of $P$-wave levels is equal to 7\%  of the $S$-state
cross-section.
\begin{table}[t]
\caption{The total cross-sections for the hadronic production of
different spin states of the $B_c$ mesons at the LHC energies.}
\begin{center}
\begin{tabular}{|c|c|c|}\hline
$nL_J$ & $\sigma$(8 TeV), nb & $\sigma$(16 TeV), nb \\ \hline
$1S_0$ & 21.961 & 46.104 \\ 
$1S_1$ & 55.039 & 115.493 \\
$2S_0$ & 4.818 & 10.115 \\
$2S_1$ & 12.022 & 25.226 \\ \hline
\end{tabular}
\end{center}
\end{table}

In fig. 2 the $d\sigma/dp_T$ distributions over the transverse momentum 
for the yields of $1S$-states of $B_c$ and $B_c^*$  are shown at the 
8 and 16 TeV energies. The maximum in the distribution is reached at
$p_T\sim M_{B_c}$.
From the distribution over rapidity (fig. 3) one can see that the
production of $B_c$ meson takes place basically in the central region.
The topology of events with the $B_c$-meson production somewhat differs 
from the configuration for the $B$ meson.
In the system of mass centre for the final $B_c+b+\bar c$ state,
the $B_c$ and $b$ particles mainly move in opposite directions.
Otherwise, as it is shown in fig. 4, the $\bar c$-quark moves
in the same direction as the $B_c$ meson. The distribution over the
angle between the $B_c$ meson and $\bar c$-quark has the sharp maximum in the
region of $0.95<\cos\theta<1$.
\begin{figure}[t]
\hspace*{1cm}$d\sigma/dy$, nb \hspace*{5cm}~$d\sigma/dy$,
nb\\
\vspace*{-19mm}
\begin{center}
\hspace*{-10mm}
\epsfxsize=12cm \epsfbox{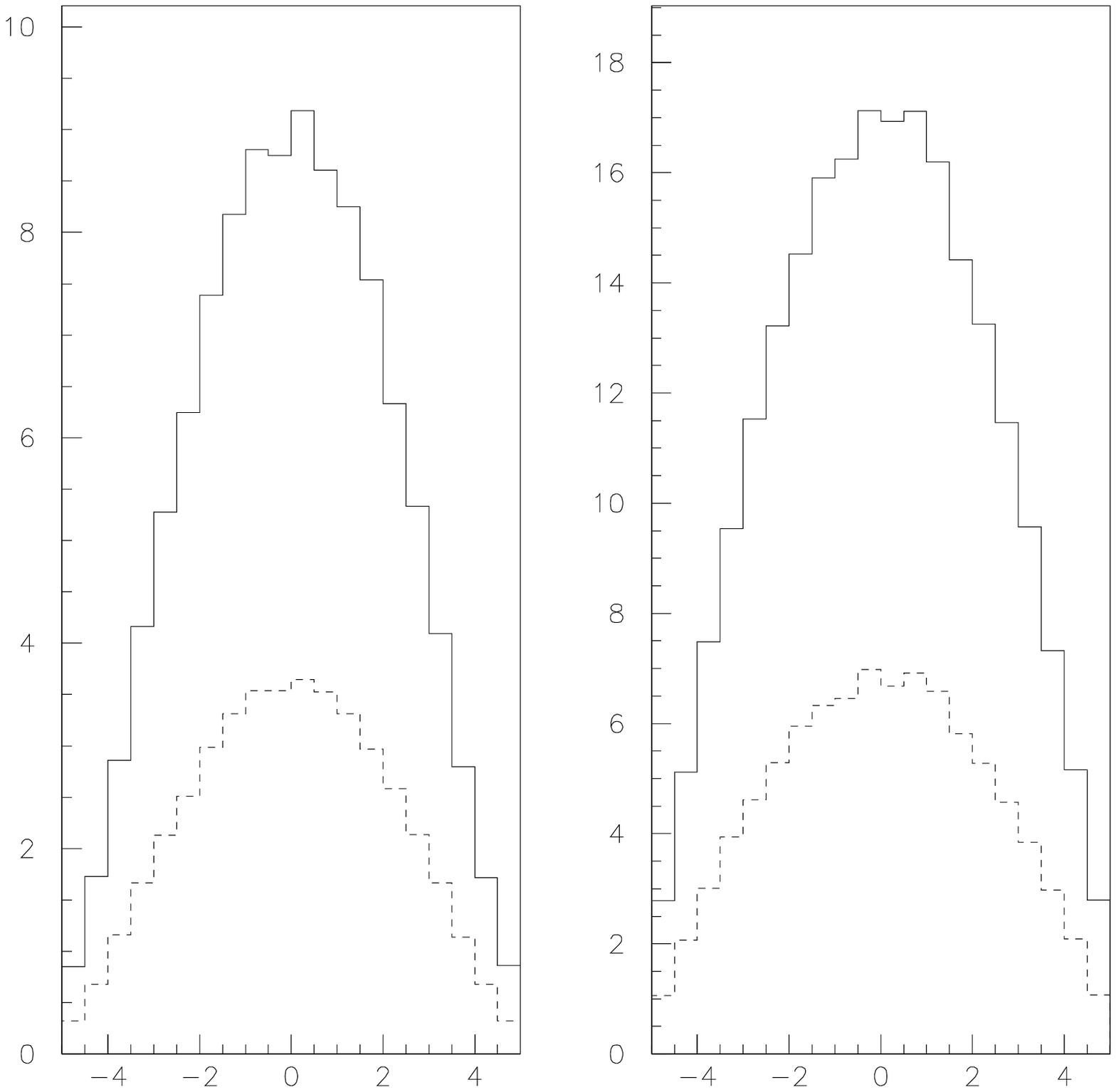}
\end{center}
\vspace*{-29mm}
\hspace*{4cm}a)~~~~~~~~~~~~~~y~~~~~~~~~~~~~~~~~~~~~~~b)~~~~~~~~~~~~~~y
\caption{The distributions over the rapidities of $B_c^*$ (solid line) and
$B_c$ (dashed line) in proton-proton interactions at the
energies of 8 TeV (a) and 16 TeV (b).}
\end{figure}

Thus, the production of $B_c$ meson is associated by the production
of $D$ meson moving in the same direction as $B_c$, i.e. the $D$ velocity
is almost parallel to the $B_c$ one, $\theta<18^0$. 
The average transverse momentum of $D$ meson is less than that of 
$B_c$ (see fig. 5) in this event .
\begin{figure}[t]
\hspace*{3cm}$d\sigma/d\cos\theta$, nb\\
\vspace*{-15mm}
\begin{center}
\hspace*{-10mm}
\epsfxsize=9cm \epsfbox{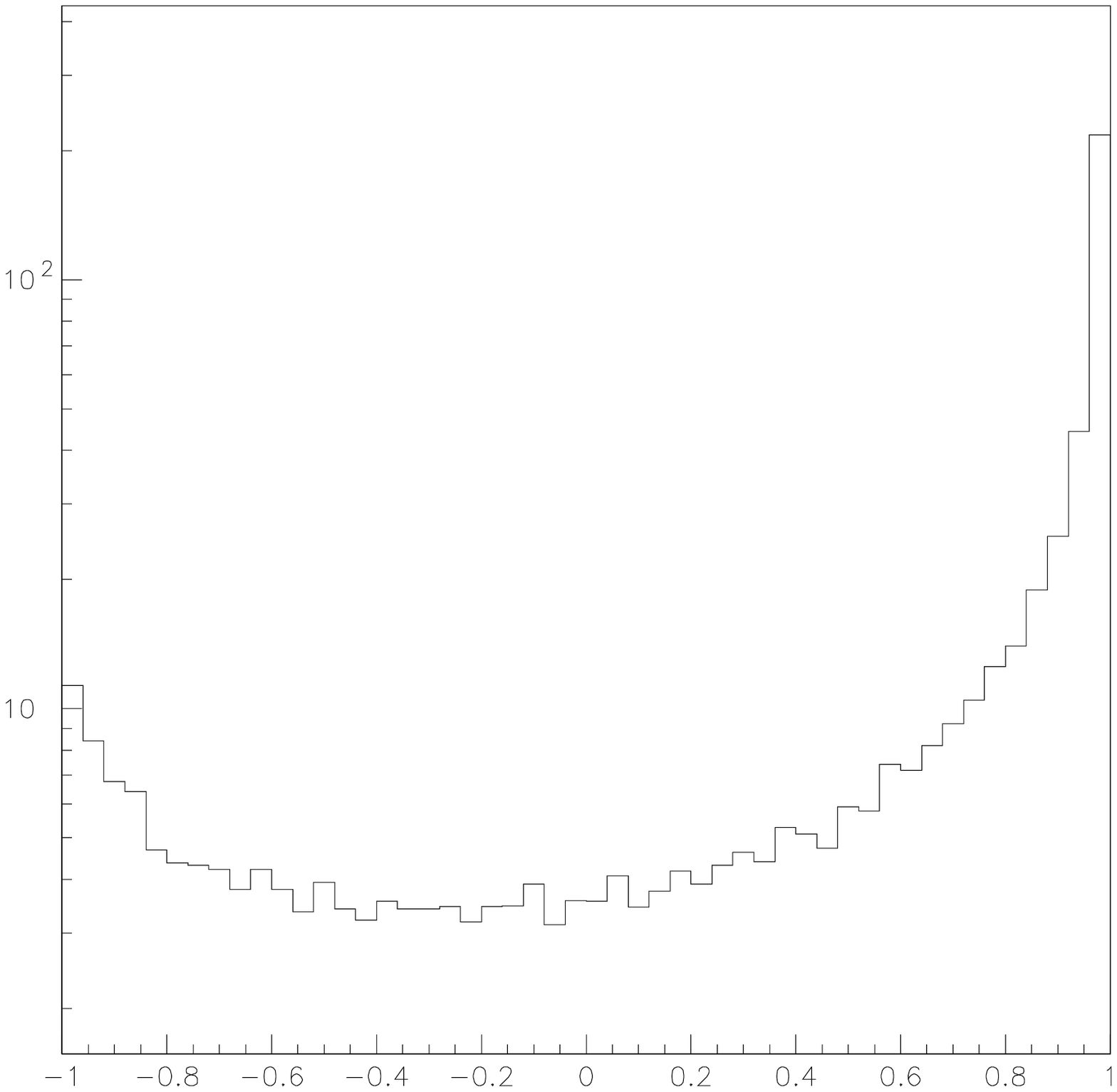}
\end{center}
\vspace*{-20mm}
{\hfill $cos\theta$\hspace*{38mm}}
\caption{The $d\sigma/d\cos\theta$ distribution over the angle of $\theta$ 
between the $B_c$ meson and $\bar c$-quark in the system of mass centre for
the colliding proton beams at the energy of 8 TeV.}
\end{figure}
As the registration of $B_c$ meson supposes the observation
of secondary vertex, two vertices must be searched for in the same jet
in the events with $B_c$. These are the vertices from the $B_c$ and 
$D$ meson decays. The correlation between the energy of $B_c$ meson and
$c$-quark associated with him is shown in fig. 6. 
A rough estimate shows that
the c-quark energy is approximately the same as that of $B_c$. Taking
into account the fact, that after the fragmentation the average 
$D$-meson energy
equals $\langle E_D\rangle\sim 0.6 \langle E_c\rangle$, and also 
the average lifetime of $B_c$ is approximately two times less than that
of the $D^+$ meson it is easily to obtain the estimates of the 
distances between the primary interaction point and the decay vertices of 
$D$ meson and $B_c$.
$$
x_D\approx \frac{E_c}{M_D}\; 0.6 c\tau_D,\;\;\;\;\;\;
x_{B_c}\approx \frac{E_{B_c}}{M_{B_c}}\; c\tau_{B_c}\;,
$$
i.e.
$$
\frac{x_D}{x_{B_c}} \simeq 3.6\;.
$$
\begin{figure}[t]
\hspace*{3cm}$d\sigma/dp_T$, nb/GeV\\
\vspace*{-15mm}
\begin{center}
\hspace*{-10mm}
\epsfxsize=9cm \epsfbox{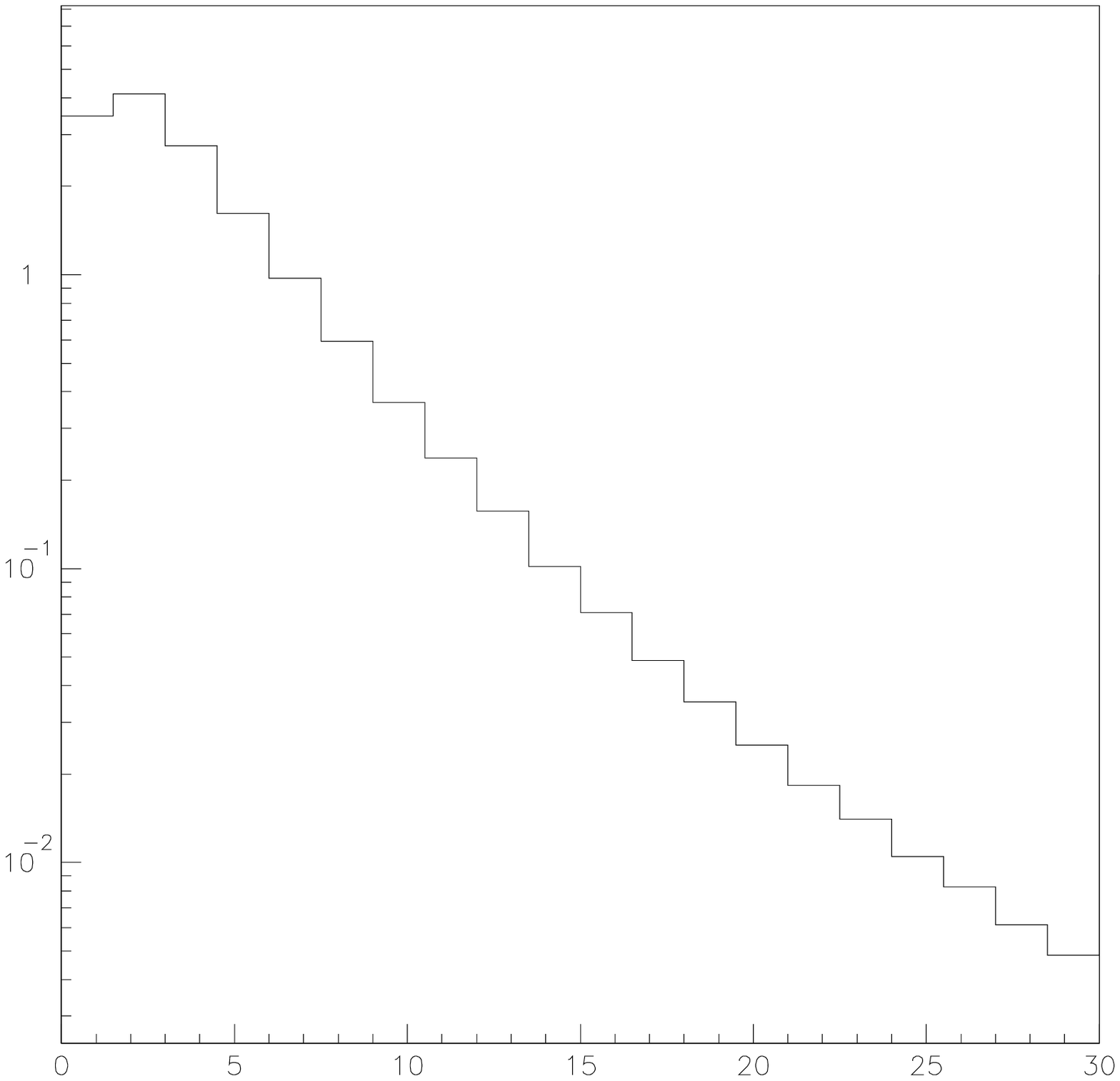}
\end{center}
\vspace*{-20mm}
{\hfill $p_T$, GeV\hspace*{38mm}}
\caption{The distribution over the transverse momentum of the $\bar c$-quark
associated with the $B_c$ meson in proton-proton interactions at the energy 
of 8 TeV.}
\end{figure}
Thus, in the events with $B_c$ one should expect also, for example, 
the associated production of $D^+$ mesons, having approximately three times
greater displacement from the primary vertex of interaction for the particles 
from the beams, than the displacement of $B_c$, and the direction of 
$D$-meson motion approximately coincides with that of $B_c$. 
This configuration can be useful in the extraction of signal from 
the hadronic background.    
\begin{figure}[t]
\hspace*{3cm}$E_{\bar c}$, GeV\\
\vspace*{-15mm}
\begin{center}
\hspace*{-10mm}
\epsfxsize=9cm \epsfbox{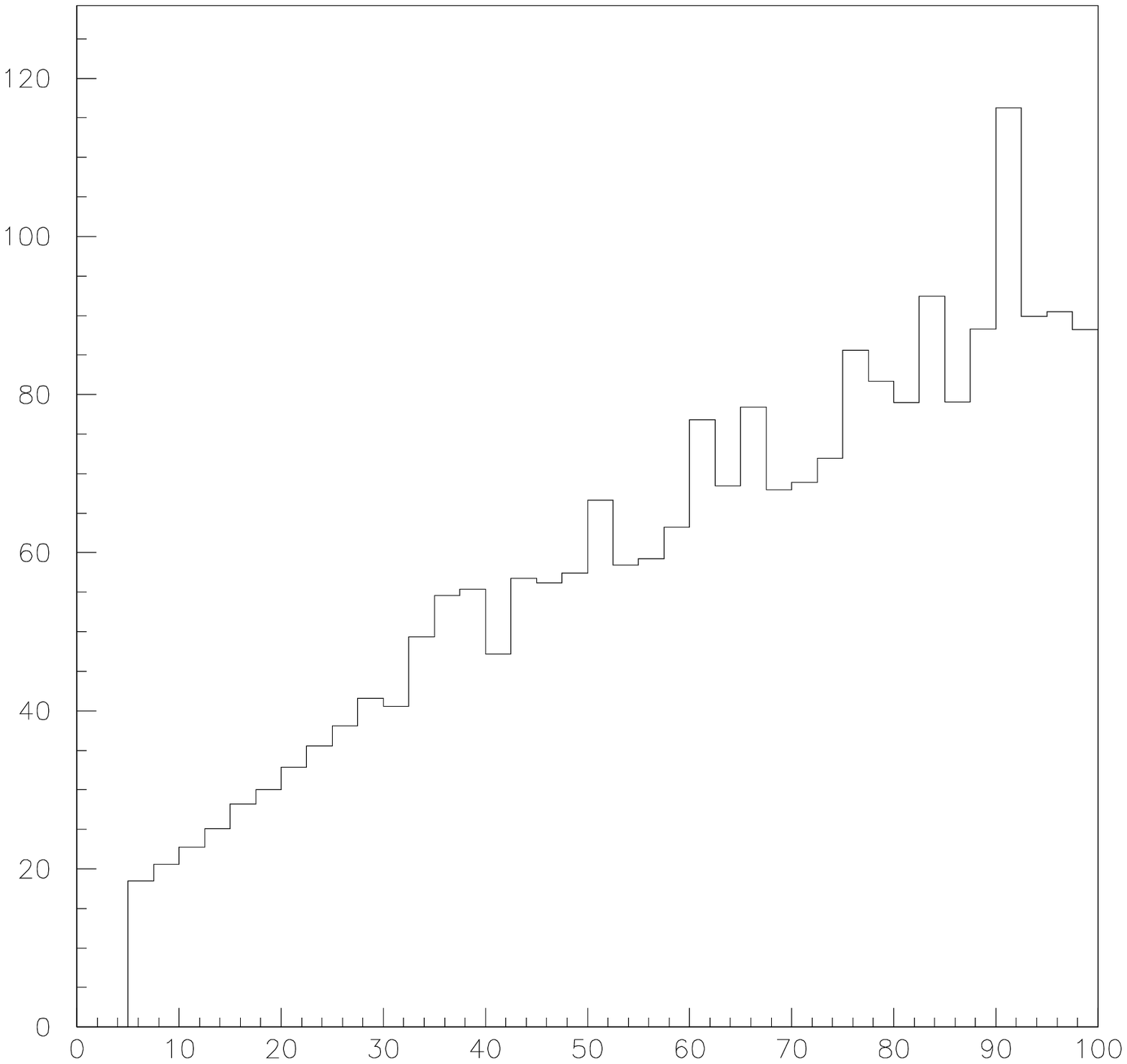}
\end{center}
\vspace*{-20mm}
{\hfill $E_{B_c}$, GeV\hspace*{38mm}}
\caption{The correlation between the energies of the $B_c$ meson and 
$\bar c$-quark associated with the former at the LHC energy of 8 TeV.}
\end{figure}
\section*{Conclusion}
In this paper we have considered the production mechanism for the different
spin  $S$- and $P$-wave states of the $B_c$ mesons in the hadron interactions
at the LHC energies on the basis of calculation for the complete set 
of diagrams for the leading
$O(\alpha_s^4)$-contribution of the perturbative QCD for the hard production
of heavy quarks and in the potential model of soft binding of the $\bar b $ 
and $c$ quarks into the meson. One has found that the regime of 
$\bar b$-quark fragmentation into the $B_c$ meson is delayed to the region
of large transverse momenta, $p_T > 6 M_{B_c}\gg M_{B_c}$, so that the 
diagrams of the nonfragmentational type, i.e. the recombinations, give
the dominant contribution into the total cross-section for the hadronic
production of $B_c$.  At the maximum luminosity 
${\cal L}=10^{34}$ cm$^{-2}$c$^{-1}$, being planned, 
one should expect the yield of 
$B_c$ at the level of $3\cdot 10^{10}$ events per year at $\sqrt{s}=8$ TeV and 
$6\cdot 10^{10}$ events per year at $\sqrt{s}=16$ TeV.

The theoretical predictions of spectroscopic characteristics for the
$B_c$-meson family and their decays allow one to carry out an object-oriented
search for the events with $B_c$ and to extract them from the hadron 
background.
For example, in the semileptonic decay mode of $B_c^+\to \psi l^+\nu_l$ 
with the account for the branching ratio for the leptonic decay  of
$\psi\to l^+l^-$ and the efficiency of the registration 
for the secondary vertex
with three leptons $\epsilon=0.1$, one should expect $12\cdot 10^6$ 
reconstructed events with the $B_c$ mesons at $\sqrt{s}=16$ TeV. Even in the
regime of low energies ($\sqrt{s}=8$ TeV) and luminosity (${\cal L}=10^{32}$
cm$^{-2}$c$^{-1}$) it is possible to reconstruct $6\cdot 10^4$ semileptonic
$B_c$ decays only in the muon channel, for example. The probability
of the $B_c^+\to \psi\pi^+$ decay has the uncertainty related with the
large momentum of the recoil $\psi$ meson. In the regime of low energies
and luminosity the number of the reliably detected events is about
$(0.4-4)\cdot 10^4$ at the same efficiency of registration of the
secondary vertex ($\epsilon=0.1$). It is worth to mention that we have pointed
out the number of events with no account for the detector cuts off the
angles and momenta in conditions of particular facilities at LHC.
Nevertheless, the search for $B_c$ at LHC seems to be the most promising
for a positive result, since, for example, at the LEP facilities
the observation of $B_c$ is extremely problematic due to the low yield 
of $B_c$. At HERA-B a strong
threshold effect for the hadron production of $B_c$ will appear because
one has to produce the additional pair of heavy $B$  and
$D$ mesons in the final state. So, the statistics at HERA-B will be 
at the same level as that of LEP, and, in addition, it will have 
conditions of a more intensive hadron background.
A low ratio of the signal to a background also does not yet allow
the CDF collaboration (FNAL) to observe the events with $B_c$ 
because of a low efficiency of the reconstruction for the secondary 
vertex \cite{cdf}. 

Thus, the search for the $B_c$ meson at LHC thinks to be a quite 
interesting and solvable problem.     

This work is in part supported by Russian Foundation for Basic Research,
grant 96-02-18216.

\hfill {\it Received January ~~, 1997}
\end{document}